**Graphene-Polyurethane Coatings for Deformable Conductors and Electromagnetic Interference Shielding**

*Pietro Cataldi[1], Dimitrios G. Papageorgiou[1,2], Gergo Pinter[1], Andrey V. Kretinin[1], William W. Sampson[1], Robert J. Young[1], Mark Bissett[1]\*, Ian A. Kinloch[1]\**

Dr. P. Cataldi, Dr. D. G. Papageorgiou, G. Pinter, Dr. A. V. Kretinin, Prof. W. W. Sampson, Prof. R. J. Young, Dr. M. Bissett, Prof. I. A. Kinloch

[1] Department of Materials, Henry Royce Institute and National Graphene Institute, University of Manchester, Oxford Road, Manchester, M13 9PL UK
[2] School of Engineering and Materials Science, Queen Mary University of London, Mile End Road, London E1 4NS, UK

E-mail: mark.bissett@manchester.ac.uk, ian.kinloch@manchester.ac.uk

Electrically conductive, polymeric materials that maintain their conductivity even when under significant mechanical deformation are needed for actuator electrodes, conformable electromagnetic shielding, stretchable tactile sensors and flexible energy storage. The challenge for these materials is that the percolated, electrically conductive networks tend to separate even at low strains, leading to significant piezoresistance. Herein, deformable conductors were fabricated by spray-coating a nitrile substrate with a graphene-elastomer solution. The coatings showed only slight increase in electrical resistance after thousands of bending cycles and repeated folding-unfolding events. The deformable conductors doubled their electrical resistance at 12% strain and were washable without changing their electrical properties. The conductivity-strain behaviour was modelled by considering the nanofiller separation upon deformation. To boost the conductivity at higher strains, the production process was adapted by stretching the nitrile substrate before spraying, after which it was released. This adaption meant that the electrical resistance doubled at 25 % strain. The electrical resistance was found sufficiently low to give a 1.9 dB/μm shielding in the 8-12 GHz electromagnetic band. The physical and electrical properties, including the EM screening, of the flexible conductors, were found to deteriorate upon cycling but could be recovered through reheating the coating.

# 1. Introduction

Electronics are ubiquitous in everyday life with the next generation of devices aiming to be fully flexible, conformable, wearable and stretchable.[1-5] Indeed, the flexible electronics market is forecasted to increase annually by 11 % and reach a market size major of 40 billion dollars by 2024.[6] One of the most established fields of stretchable electronics is strain sensing[3] which can be used in the motile parts of robots and machines[4, 7], record the movement and physiological signals in the human body[8-11] and measure mechanical deformation of solid structures[12-17]. The predominant figure of merit for strain sensors is a high gauge factor, i.e. the highest linear change in the electrical resistance as a function of deformation.[18, 19] However, numerous applications (e.g. stretchable interconnects, wearable displays, capacitive tactile sensors, printed circuit boards, deformable supercapacitors) demand stretchable conductors that maintain their electrical conductivity upon deformation and thus require a very low gauge factor, ideally zero.[20-27]

Stretchable conductors are commonly achieved by combining elastomeric matrices with conductive materials to form percolated composites. Typically the reinforcing fillers are metallic (e.g. silver nanowires/nanoflakes or copper nanowires) [28, 29] or carbon-based [21, 22, 30-32]. Polydimethylsiloxane (PDMS) is predominantly used as the matrix in research but nitriles, natural rubbers, and polyurethanes are also common.[2, 8, 20, 33-35] Alternatively, the application of conductive materials (e.g. graphene) on a pure elastomer substrate is employed.[23, 28, 31, 36] Recently, the combination of these two techniques, i.e. the application of a stretchable conductive nanocomposite coating on an elastomer, has been proposed for stretchable electrodes.[20, 32] Amongst the conductive carbon-based materials, graphene nanoplatelets (GNPs) and multiwalled carbon nanotubes (MWNTs) are available in the industrial scale at a moderate price and usually are effective at lower weight concentration levels compared to the metallic nanoparticles. Carbon nanotube-based deformable electrodes have shown promising results[21, 22, 30, 31, 37, 38], whereas graphene related materials have been thoroughly investigated for strain sensing applications[13, 39-41], their use as flexible conductors is rare[21].

Several approaches have been used to minimize the decrease in conductivity of flexible conductors upon deformation. One option is a conductive coating applied on a substrate which is then patterned, encapsulated and/or applied to pre-stretched materials.[31, 34, 42] Alternatively, nanofillers can be used to form a percolated network within a bulk composite.[18, 21, 43] Thus, it is proposed that the application of a thin, stretchable conductive nanocomposite on an elastomer could exploit both these strategies to give improved performance. Another challenge for deformable conductors is the permanent damage induced by deformation[2, 33] and thus the ability to heal such damage is highly desirable. A fundamental area of electronics that would benefit from the manufacturing of deformable and healable conductors is electromagnetic interference (EMI) shielding[43-52] given that EMI shielding is typically achieved with rigid metals. [46]

Herein, we have produced electrodes with sheet resistances of ≈ 10 Ω sq$^{-1}$ by spray coating a nitrile rubber substrate with elastomeric, conductive solutions. The sprayed solutions comprised GNPs and a thermoplastic polyurethane (TPU) binder. The final conductor thus consisted of a layered structure with a thin stretchable conductive nanocomposite on the elastomer substrate. The resistance of the GNP-TPU conductors was found to slightly change after repeated bending and folding cycles. The electrodes were washable without any detectable change in their electrical properties and doubled their initial electrical resistance at 12 % elongation. A simple semi-empirical model of the dependence of the electrical resistance on strain is proposed and discussed also. Repeated stretch-release cycles produced a mechanical deterioration of the electrical properties that can be restored through a simple heating treatment. This healing procedure restored the electromagnetic interference shielding efficiency also. Finally, pre-stretching the rubber substrate before spraying the conductive ink enhanced the electrical resistive stability of the electrodes.

## 2. Results and Discussion

2.1 Spray coating and film morphology

Nitrile substrates were spray-coated with solutions of TPU containing GNPs. This methodology was chosen due to its innate scalability. Graphene nanoplatets (GNPs, Avanzare AV240) were used with chloroform as the solvent. The GNPs were a conductive reinforcement grade and possessed a large flake diameter and highly graphitic nature. The lateral size of the GNPs was 17 ± 12 μm and their Raman spectrum showed a strong 2D peak and a D/G ratio of 0.38 (see Figures S1 and S2).

After spraying, the samples were heated to 170 °C for 20 seconds to remove the chloroform and soften the TPU to give a nanocomposite coating that conformed to the underlying nitrile substrate (Figure 1A). The polymers used are thermally stable up to 200 °C and the softening point of the TPU was 150 °C. [20, 53]. Loadings of 1 to 40 wt% GNPs were used relative to the mass of the TPU in final coating. The 40 wt% coatings were difficult to produce due to the high particulate content blocking the spray gun and thus were used for the initial concentration dependency studies only.

Scanning electron microscopy showed that the uncoated nitrile rubber had a wrinkled surface with macroscale roughness (Figure 1B), which should improve the mechanical lock-in of the coatings applied to it.[54] The heat-treated GNP-TPU films formed a uniform coating on top of the nitrile substrate with a well-adhered interface between the coating and the nitrile (Figure 1C, 1E and Figure S3). The nitrile substrate thickness was ≈ 100 μm (Figure 1D) and the final coatings were ≤ 10 μm in thickness, with the thickness being dependent on the concentration of GNPs used (Figure 1E, Figure S4 and Table S1).

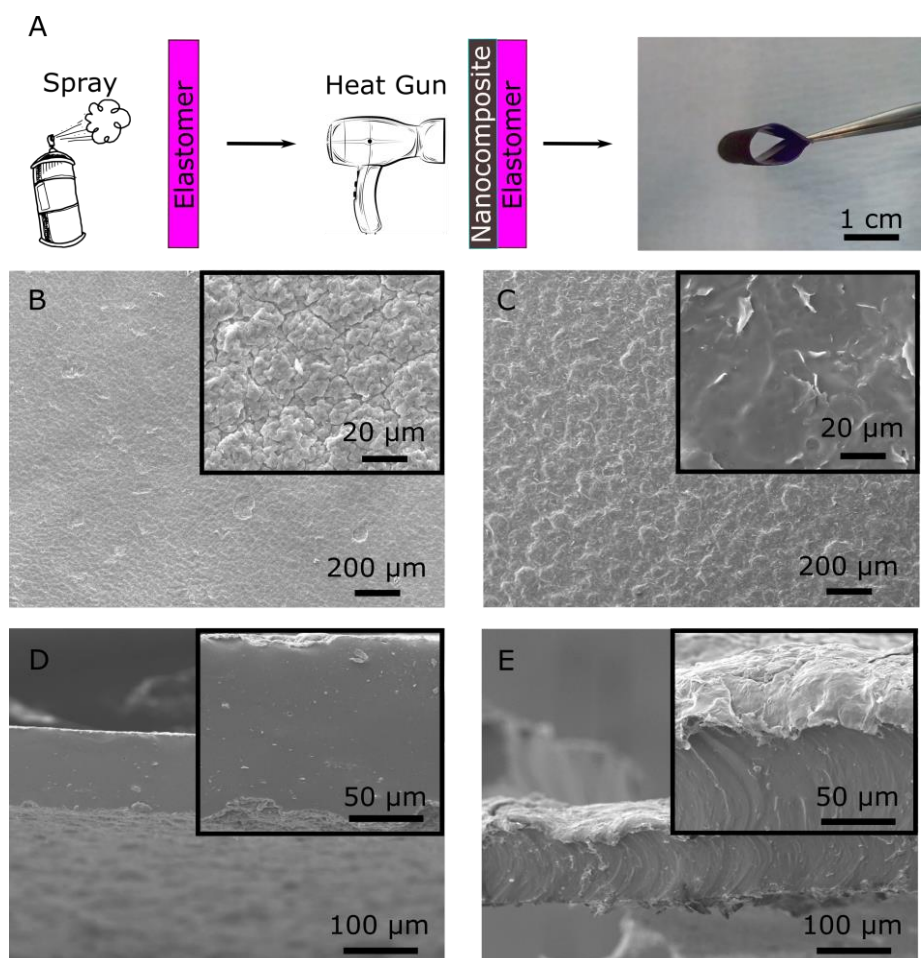

**Figure 1: A** is a schematic of the preparation of the electrode and its flexibility. TPU-GNP conductive inks were sprayed onto nitrile rubber substrates followed by heating by a heat gun. **B** and **C** show the SEM morphology of the bare nitrile rubber and of the 30 wt % GNPs concentration coating at low and high (inset) magnification, respectively. **D** and **E** show the cross-section SEM image of the same sample shown in **B** and **C**.

2.2 Electrical characterisation of the Graphene-based Coatings

The sheet resistance of the GNP-TPU was $10^{10}$ Ω sq$^{-1}$ at low GNP loading, dropped to $10^{4}$-$10^{5}$ Ω sq$^{-1}$ between 4 and 6 wt% loading and reached 9 Ω sq$^{-1}$ at the highest loading of 40 wt% GNPs (Figure 2A). This behaviour was modelled using classical percolation theory:

$$\sigma = \sigma_0 \left( \Phi - \Phi_c \right)^t \tag{1}$$

where $\sigma$ is the electrical conductivity (Table S1), $\Phi$ is the filler loading, $\Phi_c$ is the percolation threshold and $t$ is the universal critical exponent.[55, 56] Fitting to the experimental data found $\Phi_c$ as 3 wt% and $t$ as 2.9. This value of the exponent, $t$, might be expected if the GNPs formed a 3D network, however, since their diameters are greater than the film thickness, layered structures are more likely. We speculate that the high value for $t$ is may be due to significant inter-flake electron hopping in the direction perpendicular to the surface.

The key performance characteristic of flexible connectors is a minimal change in their electrical resistance upon deformation, *i.e.* as low a gauge factor as possible. This deformation may be either in axial tension or bending mode. The electrical resistance of the samples were recorded as function of applied tensile strain, $\varepsilon$ (Figure 2B). A 30 wt% loading of GNP was used in the coating (denoted "GNP-TPU" from now on) since it gave the lowest electrical resistance. Moreover, lower loading showed a higher piezoresistivity (Figure S5). Each sample was stretched up to 100 % strain at 4 % strain intervals. The current (*I*) flowing through the samples under a constant voltage of 1 V was recorded at each step (Figure 2B). The current was found to decrease with a power law behaviour, with the electrical resistance doubling at 12 % strain. At 40 % elongation the conductive elastomer preserved ~ 10 % of the initial current flow and above 80 % elongation < 0.5 µA current flowed through the samples. One reason for this increase in the electrical resistance with strain was the loss of contact between the GNP flakes inside the polymer matrix.[57] Secondly, the SEM images of the stretched coating revealed many large-scale cracks that formed on the surface of the coatings already at 30 % strain (Figure 2B and Figure S6). It should be noted that a number of applications, such as electrodes for actuators, would use strains significantly less than this 30 % strain. Upon release from 100 % strain, the deformable electrodes recovered the (25 ±

8) % of their initial conductivity. Repeated cycles at 100% strain did not alter the recovered current upon release of the strain with not further cracking occurring.

For applications, it is important to predict the dependence of the electrical resistance on $\varepsilon$. Development of a model from first principles based on the polymer physics of the composite coating is non-trivial since it would need to consider the Poisson's ratio, the interface of the GNPs and TPU, changing orientation of the GNPs, *etc*. Accordingly, such a model is beyond of the scope of this paper. Instead, we have taken a semi-empirical approach where the separation of particles due to strain was assumed to reduce the effective volume concentration of the GNPs by a factor $(1+\varepsilon)^2$. Thus, on increasing $\varepsilon$, the effective volume fraction of GNPs reduces such that the system tends towards the lower concentration of the percolation curve, increasing sheet resistance (See Figure S7). The model is more fully elaborated in the Supplementary Information and results in the relative change in conductivity at a strain $\varepsilon$ being given by

$$\frac{I}{I_0} = \left(1 - \frac{(\varepsilon-2)\varepsilon}{(1-\varepsilon)^2} \frac{\Phi}{\Phi - \Phi_c}\right)^t \qquad (2)$$

Least-squares fitting of Equation (2) to experimental data yields $\Phi_c = 2.9 \pm 0.8\%$ and $t = 2.9 \pm 0.1$ with coefficient of determination, $r^2 = 0.99$; this is shown by the solid line passing through the data in Figure 2B; we note the agreement of our estimates of $\Phi_c$ and $t$ obtained here with those arising from our percolation experiments, as shown in Figure 2A.

Flexible conductors may be used also in bending mode, for example, in polymer actuator electrode applications. Thus, the coatings were subjected to bending tests, where the electrical resistance ($R$) was measured in a flat configuration ($R_{flat}$) and when the material was curved with a 0.4 cm bending radius ($R_{bending}$). The initial flat resistance, $R_0$, of the sample was 12 $\Omega$ sq$^{-1}$. The relative change in resistance when in flat and bent morphologies, $R_{flat}/R_0$ and $R_{bending}/R_0$, upon subsequent bending cycles is shown in Figure 2C. It was found that both $R_{flat}$ and $R_{bending}$ followed the same trend: they increase for the first few cycles but then decreased upon further bending such that after ~ 50 cycles they returned to their initial values. This declining trend in $R/R_0$ was continued to the end of the test, leading to a decrease of the electrical resistance after 30,000 bending cycles of just 4% and 5 % for the $R_{flat}$ and $R_{bending}$ respectively.

The enhancement of the electrical properties after repeated mechanical deformations due to self-organization/recombination of the nanofiller inside the TPU matrix has also been reported in the literature for a polyurethane-gold[58] and polyurethane-GNP[53] nanocomposite systems. Considering practical applications, it is significant that our conductive elastomer can be subjected to tens of thousands of bending cycles without any pronounced increase of electrical resistance.

The flexible conductor was subjected to extreme bending such that would be found in a foldable device or clothing; the conductor was folded in half (i.e. 180°) and unfolded repeatedly (Figure 2D).[59] To ensure a consistent fold, the fold edge was compressed with a 1.5 kg weight on each cycle. The ratio $R_{flat}/R_0$ transverse to the fold line direction for 20 fold–unfold cycles is shown in Figure 2D. After the first cycle, the resistance increased by roughly 30 %. At the 10th fold-unfold event, $R/R_0$ reached a plateau with an increase of around the 40 % of $R_0$ which is then maintained until the 20th folding cycle. SEM revealed micrometric cracks in the coating the region of the creased created by the folding cycles which explains this change in the resistance (See Figure S8). Nevertheless, the percolating network of the nanofillers was preserved, demonstrating a remarkably low sheet resistance of 17 $\Omega$ sq$^{-1}$ after 20 folding cycles.

The mechanical properties of the nitrile rubber, pure TPU coated nitrile (TPU-Nitrile) and GNP-TPU materials were studied (Figure 2E, Figure S9, Table S2). The Young's modulus measured up to 10 % strain ($E_{10}$) of the pure nitrile was 5.7 ± 0.2 MPa. The substrate coated with a pure TPU films displayed identical mechanical behaviour within error. The GNP-TPU coating, however, made a significant difference to the mechanical behaviour despite the coating being ~10 % of the thickness of the underlying nitrile substrate. The GNP-TPU coating with 30 wt% GNP introduced an elastic region up to 20 % strain with a significantly increased $E_{10}$ of 16.1 ± 1.6 MPa. This $E_{10}$ corresponds to approximate modulus of the GNPs of 0.5 GPa, assuming that the coating and substrate could be approximated by the slab model for the rule of mixtures at such low strains and the GNPs were aligned. This modulus value for the GNPs is much lower than the 100 GPa typically quoted for GNPs. However, it is approximately within an order of magnitude to that predicted by our published model that considers the effect of the large mismatch of reinforcement and matrix modulus on the shear lag theory.[60] Interestingly, the modulus conventional used for elastomers, $E_{100}$, which is taken at 100 % strain, is similar for all the samples, as is their mechanical behaviour at higher strains. This observation, combined with the macroscale cracking observed by SEM, in the stretching samples suggests that between 20 % and 30 % strain the coating start to fails due to tearing and no longer reinforces the underlying substrate. Above 150 % strain the GNP-TPU coating follows the stress-strain curve of the bare nitrile due to the coating failing.

Finally, a conductor that maintains its original electrical resistance after numerous laundry cycles is of paramount importance for wearable devices.[61, 62] Therefore, the elastomeric conductor was laundered in water-detergent solution under stirring at 40°C for 1 hour.[62, 63] Before and after each washing cycle the water-detergent solution was replaced in order to guarantee that each cycle was performed with a suitable high surfactant concentration. As can be seen in Figure 2F, the ratio $R/R_0$ was constant during 10 washing cycles. The high washing stability of the conductor is a result of the excellent adhesion between TPU and nitrile rubber. This result is better compared with textiles functionalized with TPU-GNPs inks.[63]

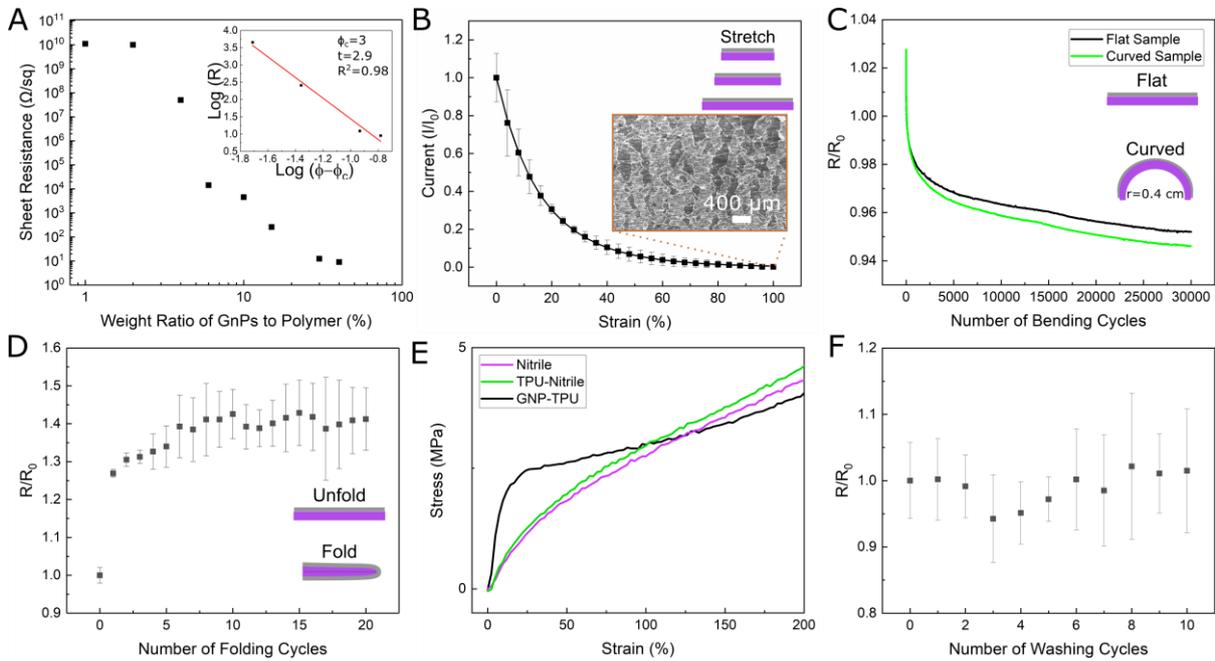

**Figure 2: A** shows the sheet resistance of the conductive elastomer as a function of the GNPs to polymer ratio. In the inset, the universal critical exponent is determined. **B** Stretch tests performed on the GNPs-based conductors and correspondent current flowing under constant voltage. The SEM image shows the sample loaded with 30 wt. % of GNPs under 100 % elongation. The black line is the fit of the model (eq. 2) to the experimental data. **C** and **D** display the bending and the folding stability of the electrical resistance, respectively, with schematics of the measuring configurations. **E** stress-strain curves of the pure nitrile, of the TPU on top of the rubber (TPU-Nitrile) and of the GNP-TPU sample. **F** washing stability of the electrical resistance of the conductive elastomer.

2.4 EMI shielding of the Deformable Conductors

Flexible and stretchable EMI shielding materials are fundamental for the implementation of conductive and flexible devices.[44-47, 64-66] Indeed, a large-scale diffusion of flexible electronics is impossible if an electronics apparatus cannot maintain a negligible interference between its electrical components when deformed.[45, 46] For this reason, the EMI shielding effectiveness of the conductive elastomer was tested depending on the amount of nanofillers employed (Figure 3), and before and after repeated stretch-release cycles (Figure 5).

The EMI shielding measurements were conducted at frequencies between 8 and 12 GHz (X-band) on transmittance. These frequencies are used in smartphones, televisions and microwaves.[67-69] The EMI shielding effectiveness (*SE*) represents the losses in the incoming electromagnetic wave due to screening and is usually calculated using:

$$SE \text{ (dB)} = -10 \log_{10}(T) \tag{3}$$

where *T* is the transmittance and represents the ratio between the transmitted and the incident electromagnetic power and is a function of the frequency of the incoming EM wave.[70] In Figure 3, the transmittance is plotted as a function of the frequency of the incident EM waves. At low nanofiller loadings (minor of 5 wt%), the samples did not display any significant electromagnetic shielding effect (0 dB attenuation, 4 wt% loaded sample plotted as an example). Increasing the filler loading, the *SE* increased with increasing nanofiller loads such that the 30 wt% GNP-TPU samples exhibited a transmittance of approximately -17.2 dB. Normalised for thickness (Table S1), the best GNPs-based sample screened 1.9 dB/μm. These normalised results are comparable with state-of-the-art attenuation levels (in the order of ≈ 1 dB/μm) of other carbon-based nanocomposites.[67, 71]

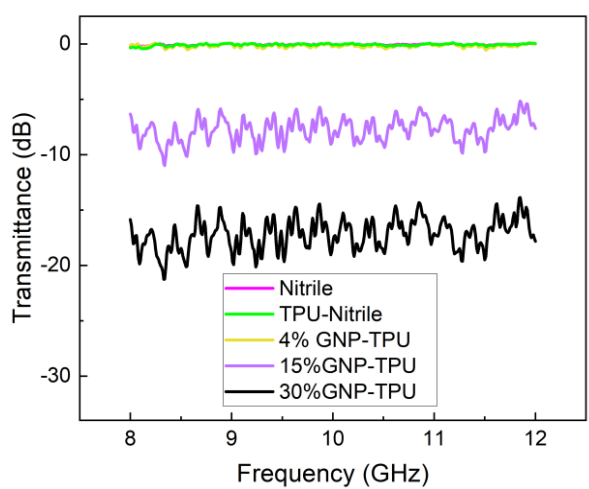

**Figure 3**: Transmittance of the conductive elastomers between 8 and 12 GHz. The percentage in the graph legend correspond the amount of nanofiller of the samples.

2.5 Healing Mechanism of the Conductive Elastomers

As discussed earlier, the conformable conductors showed hysteresis in their conductivity when deformed significantly. It was postulated that this damage could be healed through heating and re-softening the TPU, allowing the tears to heal and the percolated network to readjust. This concept was explored by applying a 1 V bias to the conductive GNP-TPU samples and then elongating them to 50 % strain. This strain would be appropriate for stretchable electronics applications such as tactile sensors and deformable printed circuit boards.[20] The force on the sample was then released and a heat gun was used to heat and re-soften the TPU coating at 170 °C. This deformation and healing process was cycled 4 times. The initial current ($I_0$), current at 50 % strain ($I_{50}$), current after each release before heat treatment ($I_{BTO}$) and current after the heat treatment ($I_{Heat}$) were measured (Figures 4A). $I_{50}$ was < 10% of $I_0$ for the samples and upon release ($I_{BTO}$) the conductivity returned to 30 – 40 % of $I_0$, showing that permanent damage had occurred. As anticipated, the heat treatment restored the conductivity of the samples ($I_H$ in the figure) of the samples to values above $I_0$. Examination in the SEM confirmed that the heating had healed the cracking within the GNP-TPU samples.

The mechanical properties of the samples at each step of the stretch-release cycles were also tested assuming the crosshead position to measure displacement. The samples were strained to 50 % elongation with their stress-strain curves given in Figure 4B. The crosshead was then returned to its initial position at which point the samples buckled slightly due to the permanent deformation that occurred during the test. The stress strain curve was then retaken (named GNP-TPU No Heat). In this last test, the stress did not immediately rise at the start of the test due to slack that had occurred from the permanent deformation. The samples were then heated using a heat gun ("Healed" in the figure) and re-stretched to 50 % elongation. The Young's modulus measured at 10 % strain ($E_{10}$) of the GNP-TPU samples was influenced by the healing treatment (Figure S10). It was found that the heating partially restored the Young's modulus for the GNP-TPU samples, taking $E/E_0$ from ≈ 0.45 to ≈ 0.70-0.80 after healing.

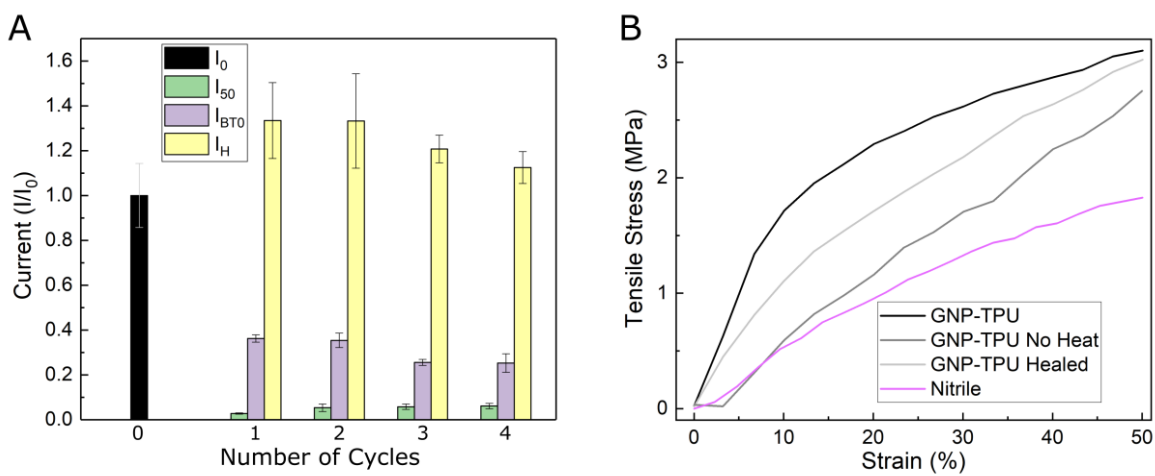

**Figure 4**: **A** displays the measurements of the current under repeated stretch release cycles at 50% stretch for the GNP-TPU samples with 30 wt% GNP. $I_{BT0}$ is the current in the samples with the stretch back to zero. After each stretch-release cycle a heating process performed with a simple heat gun permits to heal the performance loss. **B** shows the stress-strain curves till 50% elongation of the 30 wt% sample at different conditions: after fabrication (GNP-TPU), after 50% elongation without the heating treatment (GNP-TPU No Heal) and of the GNP-TPU samples after healing (GNP-TPU Healed).

Stretch-release cycles also had an influence on the EMI shielding of the conductive elastomer. The transmittance of the conductive elastomer was tested before and after repeated 100% elongation (Figure 5, for 50% elongation see Figure S11). The initial transmittance for the GNP-TPU sample ($T_0$), its value after one stretch-release cycle (GNP-TPU-Cycle1) and the value after nine stretch release cycles (GNP-TPU-Cycle9) are reported in Figure 5A. The repeated strain reduced the initial shielding effectiveness. The samples preserved ~ 75% of $T_0$ after the 9$^{th}$ strain cycle (Figure S12). Considering the thickness (see Table S1), the GNP-TPU-Cycle9 sample was able to screen ≈ 1.4 dB/μm. Even after repeated stretch-release cycles, the results are comparable with state of the art attenuation levels (of order 1 dB/μm) of other carbon-based nanocomposites.[67, 71] The stretch-induced reductions of the transmittance were once again healed by a simple heat gun procedure identical to the fabrication process (Figure 5B). The increase in the transmittance with the heat gun treatment agrees with the similar increase seen in conductivity.

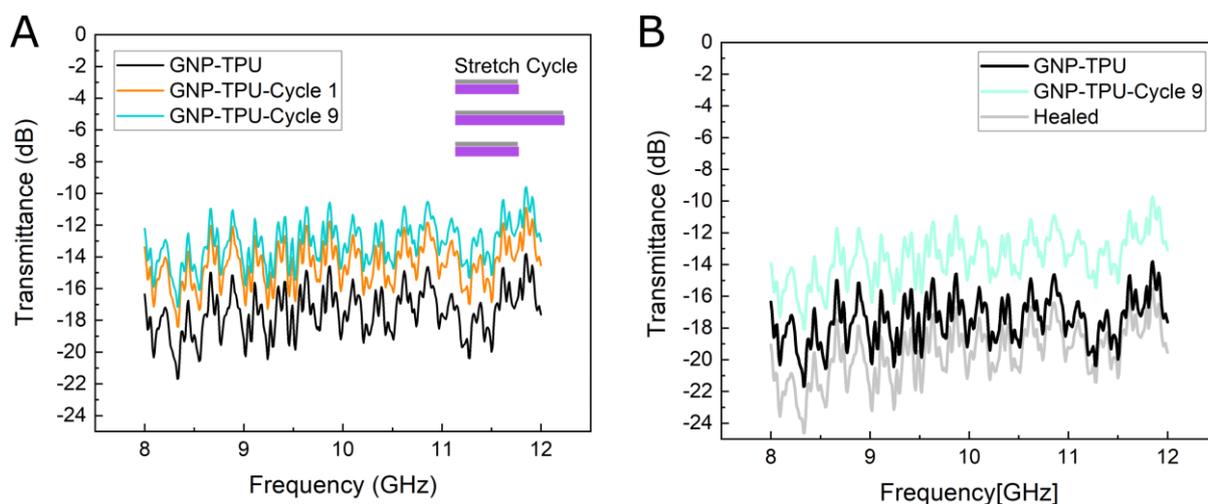

**Figure 5**: **A** Transmittance of the GNP-TPU sample before and after repeated stretch-release cycles at 100% elongation. The transmittance after one stretch-release cycle (GNP-TPU-Cycle1) and the value after nine cycles (GNP-TPU-Cycle9) are reported. **B** Healing of the transmittance with a simple heat gun procedure analogous to the fabrication process for the GNP-TPU samples.

2.6 Pre-stretching of the Rubber to Improve the Stretching Stability

The pre-stretching of elastomeric substrates before the application of a conductive layer can enhance the stability of the electrical performance of the obtained material.[31] Thus, the nitrile rubber was biaxially stretched by 50 % and then spray-coated. The biaxial strain was then released and the coating heat treated to anneal it (Figure 6A). The coating formulation studied contained 30 wt% of GNPs (PRE-GNP-TPU). An optical microscope image of the final conductor showed that the coating was very compliant (see Figure S13).

The pre-stretch production process was found not to affect the initial resistance of the samples but did significantly improve the conductivity under deformation (Figure 6B). In the first part of the current-stretch curve, the PRE-GNP-TPU samples exhibited the best performance. At 12 % elongation the PRE-GNP-TPU samples preserved 80 % of their initial current flow at a constant voltage. The gauge factor of the PRE-GNP-TPU sample was only 2 compared to 7 for the original GNP-TPU samples. The resistance of PRE-GNP-TPU doubled at 25 % strain. This performance is better than a bulk polyurethane-GNP composite[53] and a chemical vapour deposited graphene layer on PDMS[72] which showed an order of magnitude increase in the electrical resistance at 25 % stretch. Bu *et al.*[36] sprayed mechanically exfoliated graphene on top of PDMS and obtained an increase of the electrical resistance of 4 times at 20% strain. Comparable results were obtained by N. Li *et al.* [73] with a hybrid GNP-molybdenum disulphide coating on PDMS.

Our model for the strain dependence of conductivity was extended to account for pre-straining by assuming linear dependence for applied strains below the pre-strain and that Equation 2 applied for higher applied strains (see supporting information).

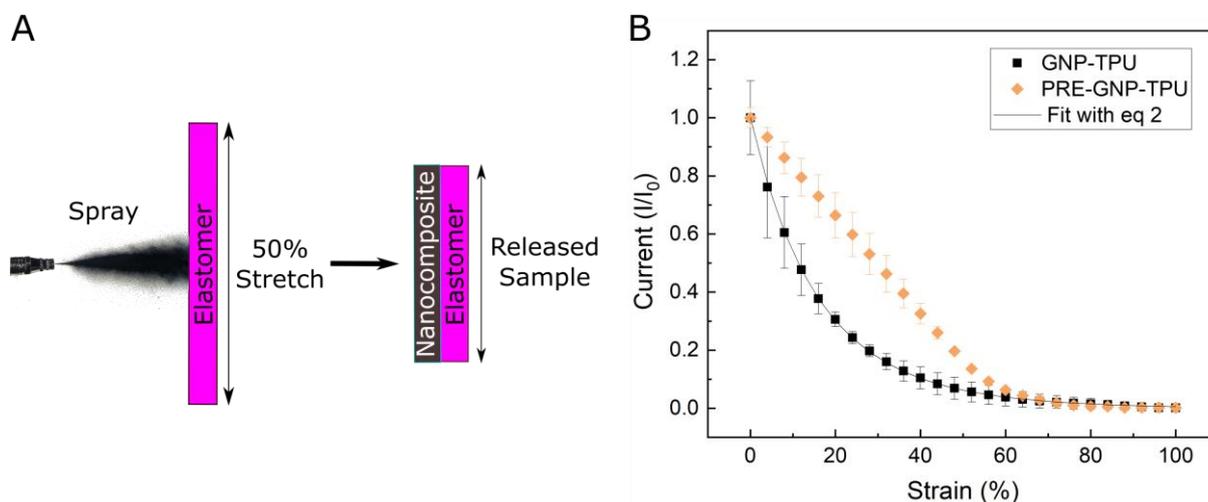

**Figure 6: A** shows a schematic of preparation of the prestretched samples. **B** Stretch test performed on the prestretched 30 wt% GNPs loaded sample (PRE-GNP-TPU) under consecutive elongation steps. The 30 wt% GNPs loaded sample (GNP-TPU) is reported for comparison.

## 3. Conclusion

Conformable conductors were produced by spray-coating nitrile substrates with an elastomeric, TPU nanocomposite containing graphene nanoplatelets. The deformable conductors had low sheet resistances of ≈10 Ω sq$^{-1}$ at 30 wt. % of nanofillers compared to the TPU. The deformable conductors maintained excellent electrical properties after thousands of bending cycles and repeated folding and washing cycles. Pre-stretching of the nitrile substrate before spraying was found to increase the stretching stability of the conductors such that their resistance doubled at 25 % strain. A simple percolation model for the relative change in conductivity for composites under mechanical deformation gives excellent agreement with experimental data. Repeated stretch-release cycles produced a mechanical deterioration that could be restored simply through heat treatment. The healing procedure also restored the electromagnetic interference shielding efficiency that was slightly reduced after repeated strain-release cycles. The realization of deformable conductors with a low change in electrical resistance upon deformation and with healable electrical features could enable the realization of truly deformable tactile sensors, actuators, wearable screens, and printed circuit boards.

## 4. Experimental Section

*Materials*: GNPs were obtained from Avanzare (grade AVA240) and were fully characterized (lateral size and Raman spectrum of the nanoflakes are presented in the supporting information, Figure S1 and S2, respectively). TPU was purchased from BASF (Elastollan 1185A12) and was used as polymer matrix. Nitrile rubber (Acrylonitrile Butadiene) were obtained from Kimberly-Clark. Typically, the conductive polymeric solution contained 0.2 g of dry TPU and a certain percentage of GNPs, indicated throughout the text as wt% ratio relative to the amount of polymer. For example, a conductive elastomer containing 30 wt% GNPs translates into a polymeric slurry having 0.06 g of GNPs. The solvent employed was chloroform (16 mL for every 0.2 g of dry TPU) acquired from Sigma Aldrich. The conductive solution was sonicated (750 W, 20 kHz, 40% amplitude, 6 times for 30 seconds, Model Num. VCX750) to achieve an adequate dispersion. After that, 4.5 mL of dispersion were spray coated (2.0 bar, 15–18 cm distance) on the rubber substrate (7.5 × 5) cm$^2$. A heat gun was employed (≈170°C, 20–25 cm distance, 30 seconds) to ensure the complete evaporation of the solvent and to enhance the adhesion. The temperature was measured using a thermocouple. For the fabrication of the pre-stretched samples, the rubber substrate was biaxially stretched of the 50% with orthogonal clamps before the application of the coating. The clamp distance was controllable simply using a screw. The sample preparation procedure was then identical to the one described above.

*Methods*: SEM pictures of the topography and of the cross section of the samples were acquired with a Zeiss Evo50 microscope (acceleration voltage of 10 kV). For cross-sectional SEM images, the specimens were frozen in liquid nitrogen and fractured.

The electrical percolation threshold was determined using a source-meter from Keithley (model 2450) in four-probe configuration. Silver conductive paint (RS pro, product number 186-3600) was painted creating 5 mm wide contacts on the samples spaced by 5 mm.

The degradation due to repeated bending cycles was determined using a custom built assembly. During bending cycles, samples were suspended between two supports. One of the supports was fixed in place while the other could oscillate horizontally along a rail system. A pneumatic cylinder (Festo Model ADN-20-50-A-P-A) controlled by an electronically switched solenoid valve (Festo Model VUVG Metric M5 5/2) was used to induce the oscillation of the moving support. To quantify the degradation, surface resistance readings were normalised to a baseline value, determined prior to any bending cycles. Four small contacts were applied to the edge of each sample using silver conductive paint (RS pro, product number 186-3600). The baseline and all subsequent surface resistance values were obtained by sweeping a DC current (Keithley Model 6221) between two contacts and measuring the resultant voltage across the other two contacts (Keithley Models 2182A). Each time, current flowed in parallel to the direction that the oscillating sample support moved. The bending cycles were periodically paused to allow normalised surface resistance readings with the sample in a flat (released) or curved orientation (bent).

The folding stability of the electrical properties after repeated stress cycles was measured recording the resistance variation transverse to the fold mark. The source-meter employed was as described above. A weight of 1.5 kg was placed on the folding edge during the folding cycle.

Washing cycles were completed by washing the conductive elastomers in water (volume of ≈ 500 ml). Ten washing cycles of 1 hour were performed. During the cycles, a water movement was maintained employing a magnetic stirrer and keeping the temperature of the water constant at 40°C. After each cycle, a detergent (Cussons Carex Complete, ≈ 4ml) was added.[63] The sheet resistance was measured with the setup described above before and after each cycle.

The I–V curves of various samples for determination of the electrical properties changes under stretching and after healing by annealing, were measured in two probe configuration. The effect of stepwise and repeated deformation on the current of the nanocomposites was characterized by the source-meter coupled with a uniaxial testing machine (Instron 3365). The samples were clamped on the testing machine and electrodes were connected to the specimen's ends. Current was recorded applying a constant potential of 1V with and without stretch. During stepwise tests, the elongation was increased by 4 % at each step with a rate of 10 mm min$^{-1}$. At each single step, deformation was held for 20 seconds to permit the sample stabilization, and afterwards the current was measured. For the cyclic tests, each cycle was performed with an elongation of the 50% of the initial length (strain rate of 10 mm min$^{-1}$), then released back to zero strain. At the end of each cycle, a heat gun procedure identical to the manufacturing process was performed to heal the material. At each stage of the cycle, the current flowing in the specimens was recorded. The uniaxial testing machine (Instron 3365, 500 mm min$^{-1}$) was also utilized for measuring the stress-strain characteristics of the stretchable conductors.

The EMI shielding effectiveness of the specimens was recorded using a vector network analyser (Keysight N5227A) and two WR-90 (8.2-12.4 GHz) waveguides. The transmittance was recorded between 8 and 12 GHz.

Raman spectra of the GNPs were obtained using a Renishaw inVia Raman spectrometer using an excitation wavelength of 514 nm.

Optical microscopy images were taken using a VHX digital microscope from Keyence.

The measurements described in this section were performed on at least three different samples.

**Acknowledgements**

This project has received funding from the European Union's Horizon 2020 research and innovation programme under grant agreement No 785219. IAK also acknowledges the Royal Academy of Engineering and Morgan Advanced Materials for funding his Chair. The authors acknowledge Prof Thomas Thomson and Harry Waring for the support with the EMI shielding measurements. The authors also acknowledge Phillip Higgins for the help in designing the uniaxial stretch-holder for the SEM and the biaxial stretch holder.

Received: ((will be filled in by the editorial staff))
Revised: ((will be filled in by the editorial staff))
Published online: ((will be filled in by the editorial staff))

**Keywords: stretchable electronics, healable electronics, conformable electronics, thermoplastic polyurethane, piezoresistivity**

Pietro Cataldi, Dimitrios G. Papageorgiou, Gergo Pinter, Andrey V. Kretinin, William W. Sampson, Robert J. Young, Mark Bissett*, Ian A. Kinloch*

**Graphene-Polyurethane Coatings for Deformable Conductors and Electromagnetic Interference Shielding**

Electrodes with stable electronics features under mechanical deformation are the holy grail of stretchable electronics. Conformable electrodes are fabricated functionalizing rubber with elastomeric nanocomposite containing graphene. These electrodes show exceptional deformation stability. The electrical performance deterioration induced by stretch release cycles can be healed trough simple heating procedures, restoring also the electromagnetic interference shielding efficiency.

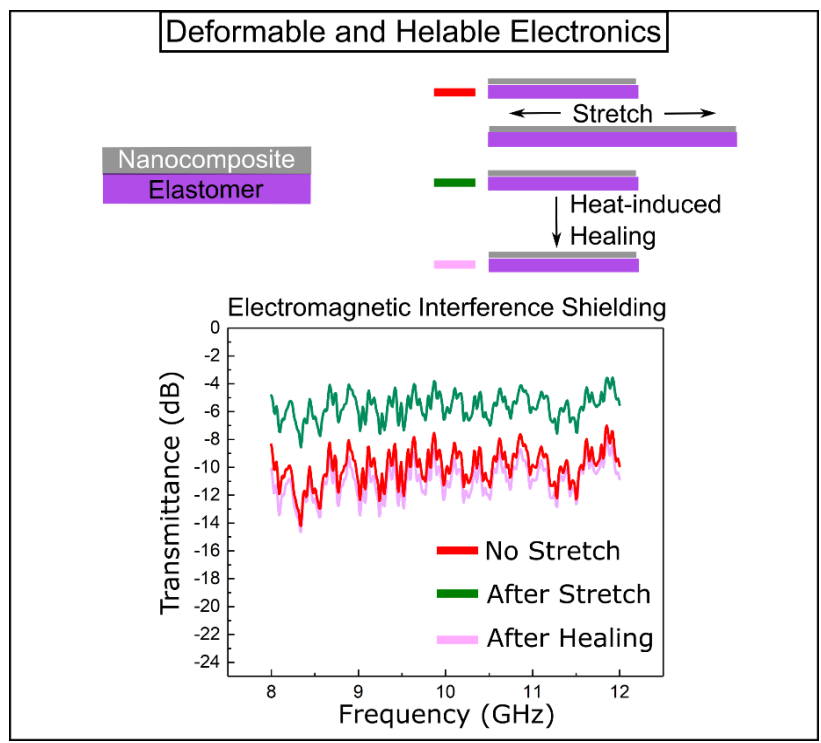

Supporting Information

**Graphene-Polyurethane Coatings for Deformable Conductors and Electromagnetic Interference Shielding**

*Pietro Cataldi[1], Dimitrios G. Papageorgiou[1,2], Gergo Pinter[1], Andrey V. Kretinin[1], William W. Sampson[1], Robert J. Young[1], Mark Bissett[1]\*, Ian A. Kinloch[1]\**

Dr. P. Cataldi, Dr. D. G. Papageorgiou, G. Pinter, Dr. A. V. Kretinin, Prof. W.W. Sampson, Prof. R. J. Young, Dr. M. Bissett, Prof. I.A. Kinloch

[1] Department of Materials and National Graphene Institute, University of Manchester, Oxford Road, Manchester, M13 9PL UK
[2] School of Engineering and Materials Science, Queen Mary University of London, Mile End Road, London E1 4NS, UK

E-mail: mark.bissett@manchester.ac.uk, ian.kinloch@manchester.ac.uk

**Graphene Nanoplatelets Lateral Size Distribution**

Figure S1 shows the lateral size distribution of the graphene nanoplatelets (GNPs). Figure S1a displays a SEM images of GNPs after tip sonication and spraying on top of a silicon substrate. The tip sonication and spray was performed as described in the method section of the main text. The concentration of the GNPs dispersion was 0.2 mg/ml. From SEM images, the AVA240 GNPs showed a broad distribution of the lateral size after tip sonication and spray, with most of the nanoflakes displaying sizes between 5 and 30 μm.

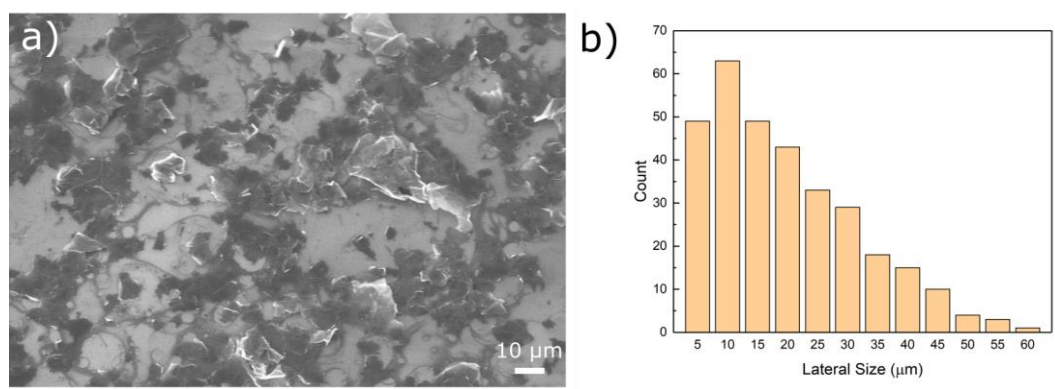

**Figure S1**: a) Representative SEM image of the nanoflakes after tip sonication and spray. b) lateral size distribution of the nanoflakes extracted by the SEM images.

**Raman Spectroscopy**

Figure S2 displays the RAMAN spectra of the AVA240 nanoflakes after tip sonication and spray coating.

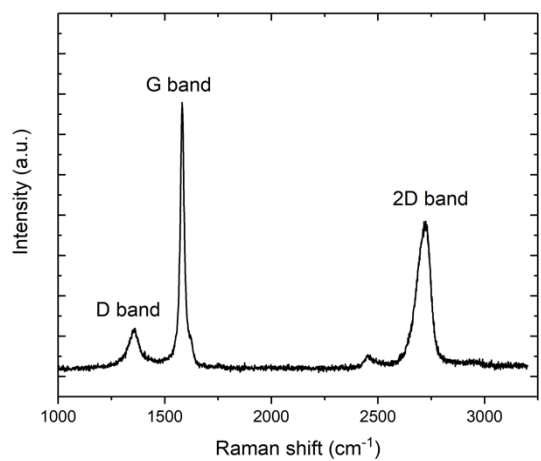

**Figure S2:** Raman spectrum of the graphene nanoplatelets

**Scanning Electron Microscopy: Morphology**

Figure S3 displays the SEM morphologies of the samples. Figure S3a and Figure S3b shows the specimens with 1 and 10 wt% ratio of GNPs to thermoplastic polyurethane (TPU) polymer, respectively**.**

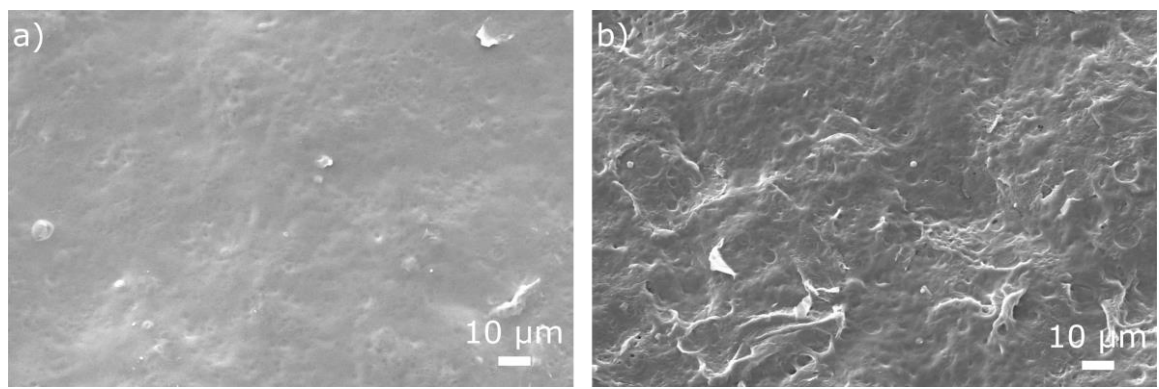

**Figure S3:** a) and b) SEM topographies of specimens with 1 and 10 wt. % ratio of GNPs to TPU polymer sprayed on top of the nitrile substrate, respectively.

**Scanning Electron Microscopy: Cross section**

Figure S5 displays the SEM cross sections of the samples. Figure S4a and Figure S4b shows the specimens with 1 and 10 wt% ratio of GNPs to thermoplastic polyurethane (TPU) polymer, respectively.

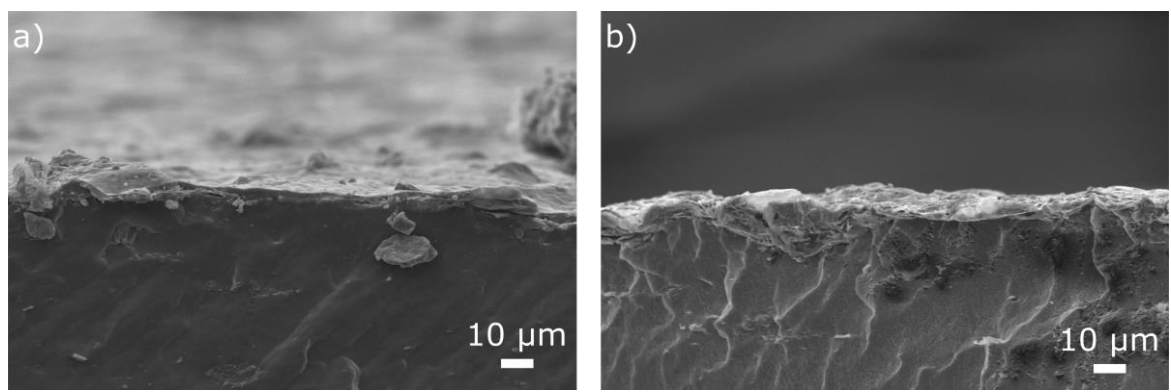

**Figure S4:** a) and b) SEM cross sections of specimens with 1 and 10 wt. % ratio of GNPs to TPU polymer sprayed on top of the nitrile substrate, respectively. The thickness of the lower concentrated GNPs material is around 4 μm while the higher concentrated has a thickness of approximately 7 μm.

**Current vs Elongation: Graphene Nanoplatelets-based Samples**

Figure S5 presents the current flowing in the graphene-based samples as a function of elongation. The measurements were performed as described in the methods section. A higher concentration of GNPs increase the strain stability.

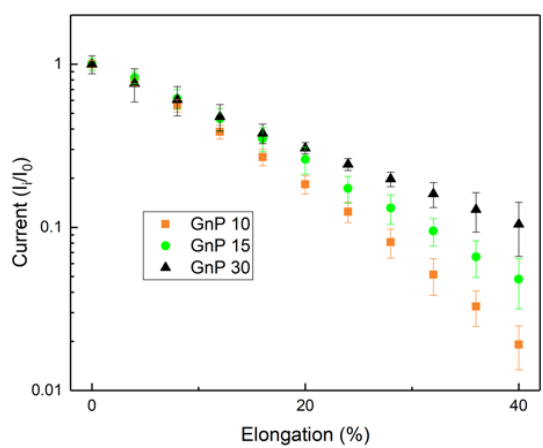

**Figure S5:** Stretch tests performed on the GNPs-based conductors and correspondent current flowing under constant voltage.

**SEM of the Cracked Sample**

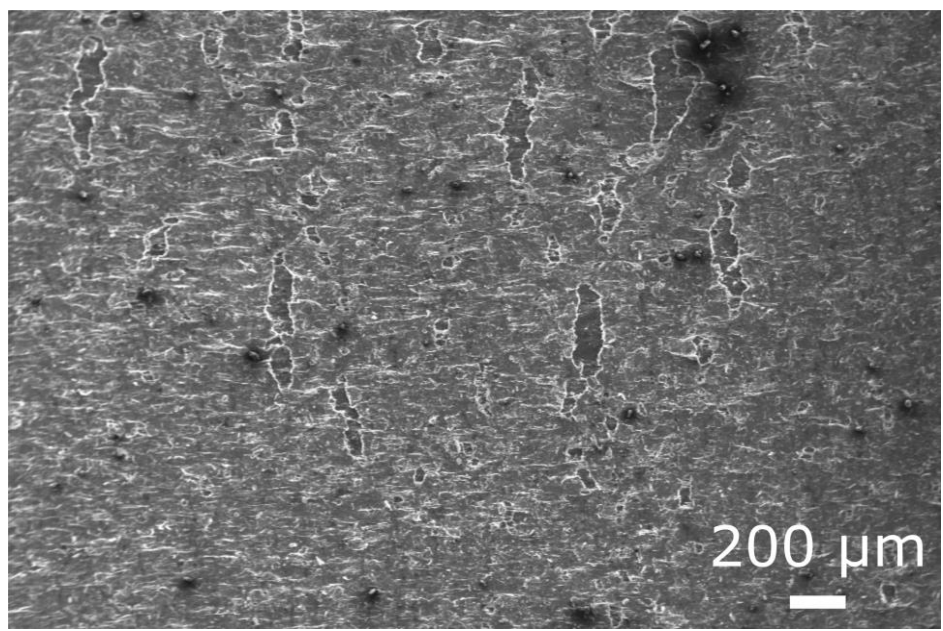

**Figure S 6:** SEM of the crack formation of the 30 wt% GNP loaded coating at 30% strain.

## Modelling Conductivity of Strained GNP Films

Assume that the conductivity of a film with solids content $\Phi > \Phi_c$, where $\Phi_c$ is the percolation concentration, is given by:

$$\sigma = \sigma_0 (\Phi - \Phi_c)^t \tag{S1}$$

If the film is deposited on a rubber substrate and subjected to a uniaxial strain, $\varepsilon$, then the particles will separate from each other in the direction of straining, effectively reducing their concentration (see Figure S8).

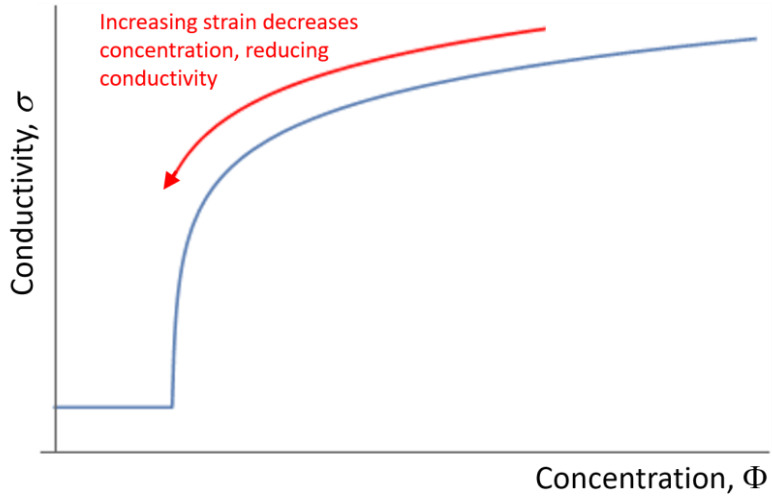

**Figure S7:** Schematic showing the equivalency of strain and reducing the effective percolation.

Perpendicular to the direction of strain, we assume that the film experiences the same Poisson contraction as the substrate and that this results in out-of-plane 'tilting' of platelets, again reducing concentration. For simplicity, we assume that to a first approximation the concentration change is the same in both directions such that $\Phi \to \Phi/(1+\varepsilon)^2$, and from Eqn. (S1) the relative change in conductivity in a network under uniaxial strain $\varepsilon$ is

$$\frac{I}{I_0} = \frac{\left(\dfrac{\Phi}{(1+\varepsilon)^2} - \Phi_c\right)^t}{(\Phi - \Phi_c)^t}$$

$$= \left(1 - \frac{(\varepsilon - 2)\varepsilon}{(1-\varepsilon)^2} \frac{\Phi}{\Phi - \Phi_c}\right)^t \quad \text{for } \Phi > (1+\varepsilon)^2 \Phi_c \tag{S2}$$

Consider now a substrate subjected to a uniaxial strained, $\varepsilon_p$, and held at this strain whilst a film is applied with solids content $\Phi > \Phi_C$ and allowed to relax. Under subsequent straining, we expect a linear dependence of conductivity on strain when $\varepsilon < \varepsilon_p$ and that Equation (2) with $\varepsilon \to (\varepsilon - \varepsilon_p)$ will hold at higher strains. Making this substitution and simplifying yields

$$\frac{I}{I_0} = \begin{cases} 1 - k\varepsilon & \text{for } \varepsilon < \varepsilon_p \\ (1 - k\varepsilon_p)\left(\dfrac{\Phi - (1+\varepsilon-\varepsilon_p)^2 \Phi_c}{(1+\varepsilon-\varepsilon_p)^2(\Phi - \Phi_c)}\right)^t & \text{for } \varepsilon \geq \varepsilon_p \end{cases} \quad (S3)$$

Note that Equation (S3) recovers Equation (S2) when $\varepsilon_p = 0$.

Table S3: Fit of equation (S3) with the experimental data.

| Parameter | GNP | GNP-Prestrained |
|---|---|---|
| t | 2.9 | 4.6 |
| k | -- | 1.66 |
| $\varepsilon_p$ | -- | 0.495 |

**SEM after Folding**

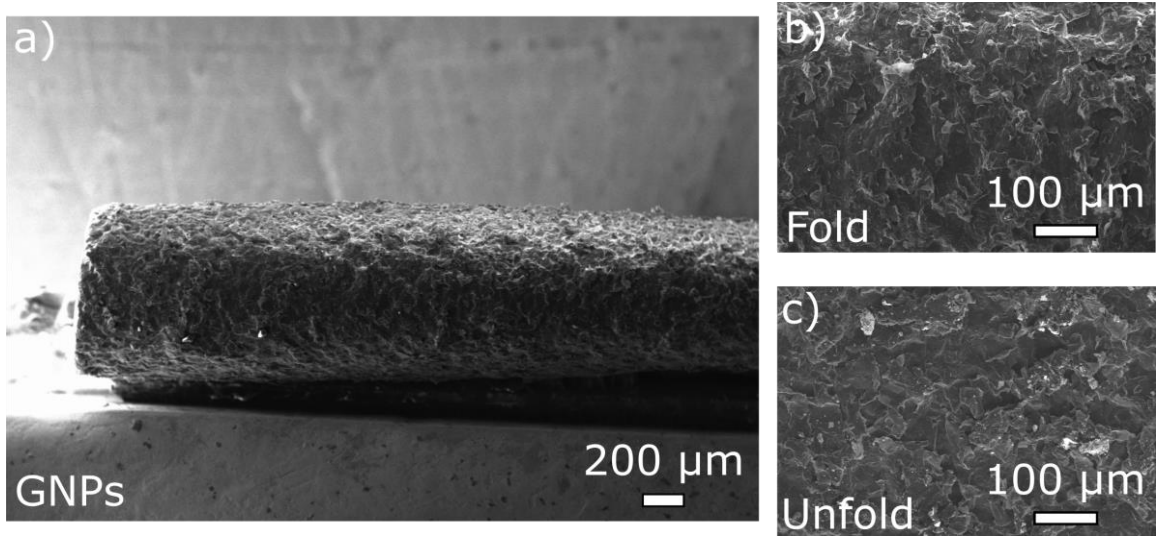

**Figure S8:** SEM images of the samples after repeated fold-unfold cycles. a)-c) are images of the graphene-based coating (30 wt% loading). They present cracks on the surfaces.

**Mechanical Properties**

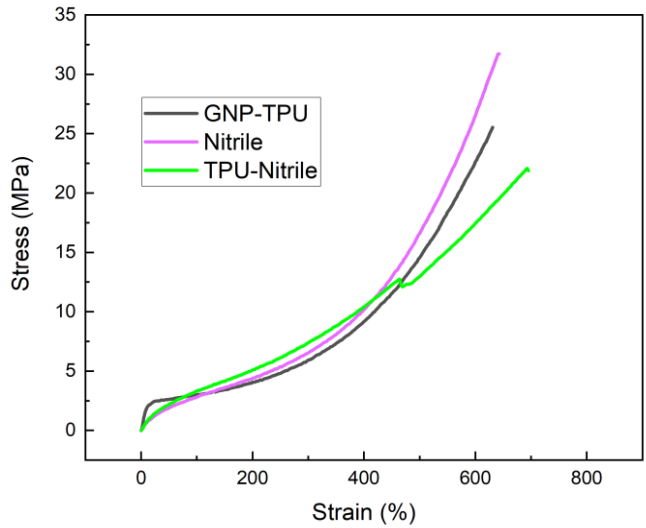

**Figure S9**: Stress strain curves of the samples. GNP-TPU is the 30 wt% loaded sample.

**Young's Modulus Healing**

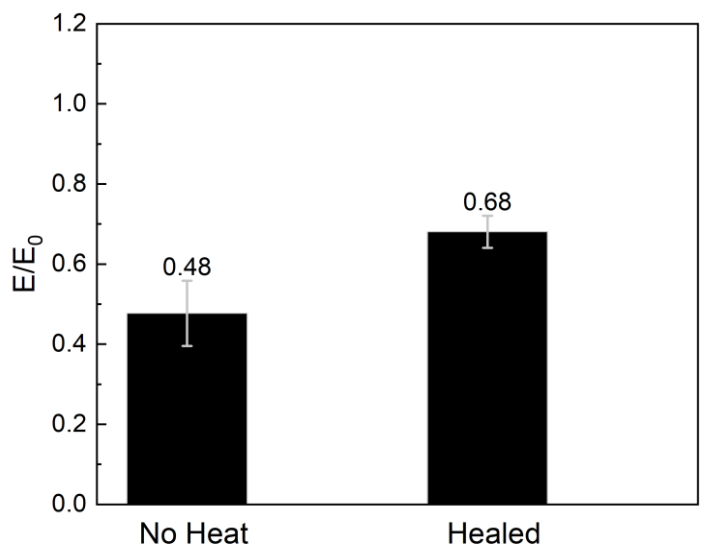

**Figure S 10**: displays the ratio $E/E_0$ between the Young Modulus before and after the heating treatment (E) and the Young Modulus after fabrication ($E_0$). $E/E_0$ change from 0.45 to 0.7-0.8 before and after the heat-assisted healing treatment, respectively.

**50% Stretch Release Cycles: Effect on EMI shielding**

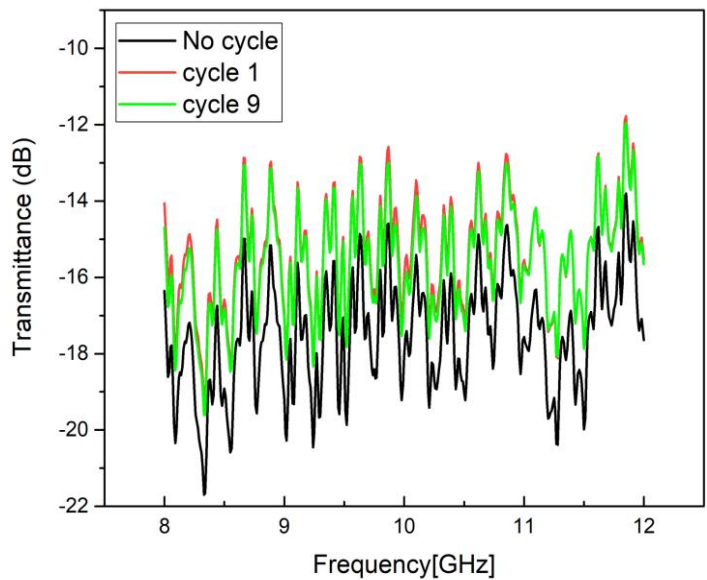

**Figure S11:** Transmittance of the 30 wt% samples after repeated stretch-release cycles at 50% elongation.

**Transmittance Change with Stretch-release Cycles**

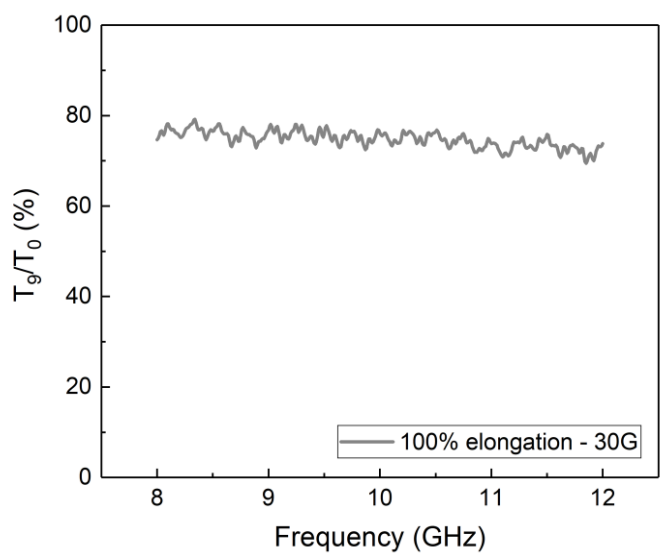

**Figure S12:** Point by point ratio between the transmittance after 9 stretch-release cycles ($T_9$) and its initial value ($T_0$) for the 30 wt% GNPs sample.

**Optical Microscope Image of the Pre-stretched Sample**

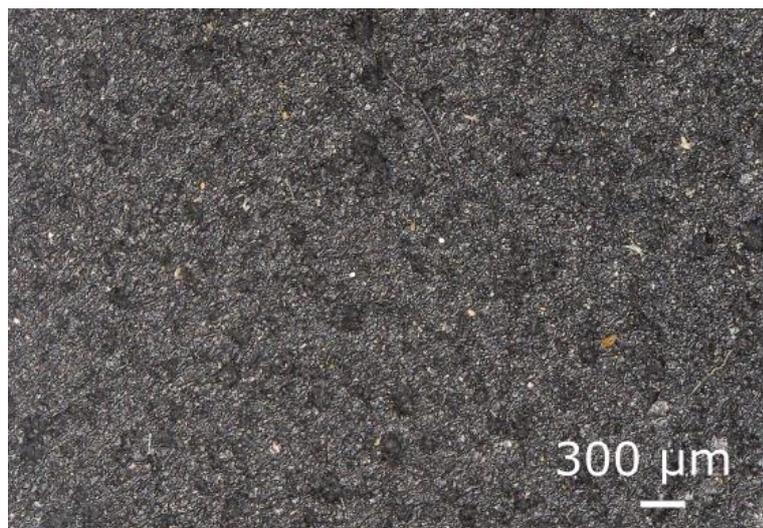

**Figure S 13**: Optical microscope image of the pre-stretched sample coating containing 30wt% GNP.

| Sample | Thickness (μm) |
|---|---|
| Nitrile Rubber | 101 ± 6 |
| TPU coating | 3 ± 3 |
| 1% GNPs Coating | 3 ± 3 |
| 10% GNPs Coating | 6 ± 3 |
| 30% GNPs Coating | 9 ± 4 |

**Table S1:** Thicknesses of the samples

| Sample | $E_{10}$ (MPa) | $\sigma_f$ (MPa) | $\varepsilon_f$ (%) |
|---|---|---|---|
| Nitrile rubber substrate | 5.7 ± 0.2 | 31.9 ± 4.6 | 624.0 ± 55.0 |
| TPU sprayed on nitrile | 6.3 ± 0.7 | 28.7 ± 6.9 | 578.8 ± 93.7 |
| 1% GNPs | 8.1 ± 0.2 | 27.1 ± 2.7 | 677.3 ± 34.0 |
| 10% GNPs | 10.6 ± 0.9 | 29.4 ± 4.2 | 665.3 ± 30.3 |
| 30% GNPs | 16.1 ± 1.6 | 26.8 ± 1.7 | 651.8 ± 32.6 |

**Table S2:** Mechanical properties of the samples. Column E indicate the Young's Modulus, column $\sigma_f$ the stress at break and column $\varepsilon_f$ the elongation at break.